\documentclass[letterpaper, 10 pt, conference]{ieeeconf}

\IEEEoverridecommandlockouts
\usepackage{graphicx}
\usepackage[hidelinks]{hyperref}
\usepackage{url}
\usepackage{amsmath,amssymb,amsfonts}
\usepackage{algorithmic}
\usepackage{textcomp}
\usepackage{xcolor}
\usepackage{multirow}
\usepackage{booktabs}
\usepackage{subcaption}
\usepackage{array}
\usepackage{tcolorbox}
\usepackage{stfloats}
\title{\LARGE \bf EEGForceFusion: Joint Tokenised--Continuous Representation Learning for Subject-Independent Grasp Force Decoding}

\author{Sankalp Sunil Turankar$^{1}$, Yogesh Kumar Meena$^{1}$%
\thanks{$^{1}$Sankalp Sunil Turankar and Yogesh Kumar Meena are with Human-AI Interaction (HAIx) Lab, IIT Gandhinagar, India {\tt\small yk.meena@iitgn.ac.in}}
}

\begin{document}
\maketitle

%%%%%%%%%%%%%%%%%%%%%%%%%%%%%%%%%%%%%%%%%%%%%%%%%%
% ABSTRACT
%%%%%%%%%%%%%%%%%%%%%%%%%%%%%%%%%%%%%%%%%%%%%%%%%%

\begin{abstract}
Brain--machine interfaces provide a link between neural activity and external devices, enabling restoration of motor function and advancing human--machine interaction using non-invasive electroencephalography (EEG). However, continuous grasp force decoding remains challenging due to complex temporal dynamics, high inter-subject variability, and limited generalisation of existing approaches. To address this, we propose a hybrid EEG decoding framework that jointly models continuous and tokenised representations, enabling capture of both fine-grained neural structure and long-range temporal dependencies. The proposed approach integrates convolutional--recurrent representation learning, quantisation-based tokenisation, and transformer-based temporal modelling within a unified fusion-based regression architecture. Experimental evaluation on the WAY-EEG-GAL dataset under strict leave-one-subject-out conditions achieves $R^2 = 0.817$ in offline settings and $R^2 = 0.793$ in simulated real-time evaluation, with latency suitable for real-time deployment. These results demonstrate strong cross-subject generalisation and highlight the practicality of hybrid continuous--tokenised representations for real-time EEG-based force decoding in assistive robotics, neuro-rehabilitation, and human--machine interaction.
\end{abstract}

%%%%%%%%%%%%%%%%%%%%%%%%%%%%%%%%%%%%%%%%%%%%%%%%%%
% INTRODUCTION
%%%%%%%%%%%%%%%%%%%%%%%%%%%%%%%%%%%%%%%%%%%%%%%%%%

\section{Introduction}

Brain--machine interfaces (BMIs) have emerged as a promising paradigm for restoring motor function and enabling assistive technologies by decoding neural activity into control signals~\cite{ng, Mane2020-ja}. In particular, non-invasive electroencephalography (EEG)-based BMIs have gained significant attention due to their safety, portability, and applicability in neuro-rehabilitation and human--machine interaction systems~\cite{hortal2015using,xie2022brain}. While substantial progress has been made in decoding kinematic variables, such as movement trajectory and velocity, accurately estimating kinetic variables, such as continuous grasp force, remains a challenging problem.

Decoding grasp force from EEG signals is inherently difficult due to complex temporal dynamics and significant inter-subject variability. Neural activity related to force generation involves distributed cortical regions, including the parietal--premotor network~\cite{ds09, Ehrsson2001-nf}, and evolves across multiple temporal scales. In addition, the low signal-to-noise ratio and presence of artefacts further complicate reliable prediction~\cite{fatourechi2007emg}.

Existing approaches for EEG-based force decoding fall broadly into feature-engineering-based and deep learning-based methods. Traditional pipelines, such as EEGForceMap~\cite{dangi2025eegforcemap}, rely on handcrafted time-frequency features followed by regression models. While effective, these approaches depend on predefined features and are limited in their ability to capture complex neural dynamics across subjects. Deep learning approaches, including convolutional and recurrent architectures~\cite{10.1007/978-981-99-8021-5_7,kumarasinghe2021brain,pgpsc19}, learn representations directly from EEG signals but struggle to jointly model fine-grained neural structure and long-range temporal dependencies. Consequently, subject-independent performance remains limited.

To address these limitations, we propose a tokenisation-based EEG decoding framework that bridges continuous and discrete representations for structured temporal modelling. Tokenised representations enable modelling of long-range temporal dependencies but lose fine-grained signal detail, whereas continuous embeddings preserve detailed neural information but lack explicit temporal structure. This trade-off motivates a unified framework that jointly models both representation types to capture local neural dynamics and long-range temporal dependencies simultaneously. 

In the proposed framework, EEG signals are first encoded into compact embeddings using a CNN-GRU encoder~\cite{bouchane2025hybrid}. These embeddings are then discretised using quantisation-based tokenisation for transformer-based sequence modelling~\cite{pradeepkumar2026tokenizing, vaswani2017attention}. Local temporal variations are captured using embedding differences, and all representations are combined within a fusion-based regression framework to predict continuous grasp force.

A key design choice is to use the transformer as a temporal feature extractor rather than a standalone predictor. Trained in a shallow regime, the transformer captures general temporal patterns while reducing overfitting to subject-specific dynamics~\cite{kostas2021bendr}, thereby improving subject-independent generalisation. The main contributions of this work are:
\begin{itemize}
    \item proposing a tokenisation-based EEG representation framework that enables transformer-based temporal modelling without relying on handcrafted features.
    \item introducing a unified architecture that combines continuous embeddings and tokenised representations for multi-scale modelling of neural dynamics.
    \item demonstrating that discrete token sequences can be fused with continuous embeddings to provide a compact, frozen temporal representation for real-time force prediction while preserving decoding accuracy.
    \item evaluating the proposed framework under strict leave-one-subject-out and simulated real-time conditions on the publicly available WAY-EEG-GAL dataset, and comparing its performance with state-of-the-art methods.
\end{itemize}

%%%%%%%%%%%%%%%%%%%%%%%%%%%%%%%%%%%%%%%%%%%%%%%%%%
% METHODS
%%%%%%%%%%%%%%%%%%%%%%%%%%%%%%%%%%%%%%%%%%%%%%%%%%

\section{Methods and Materials}

\begin{figure*}[!t]
    \centering
    \includegraphics[width=.82\textwidth]{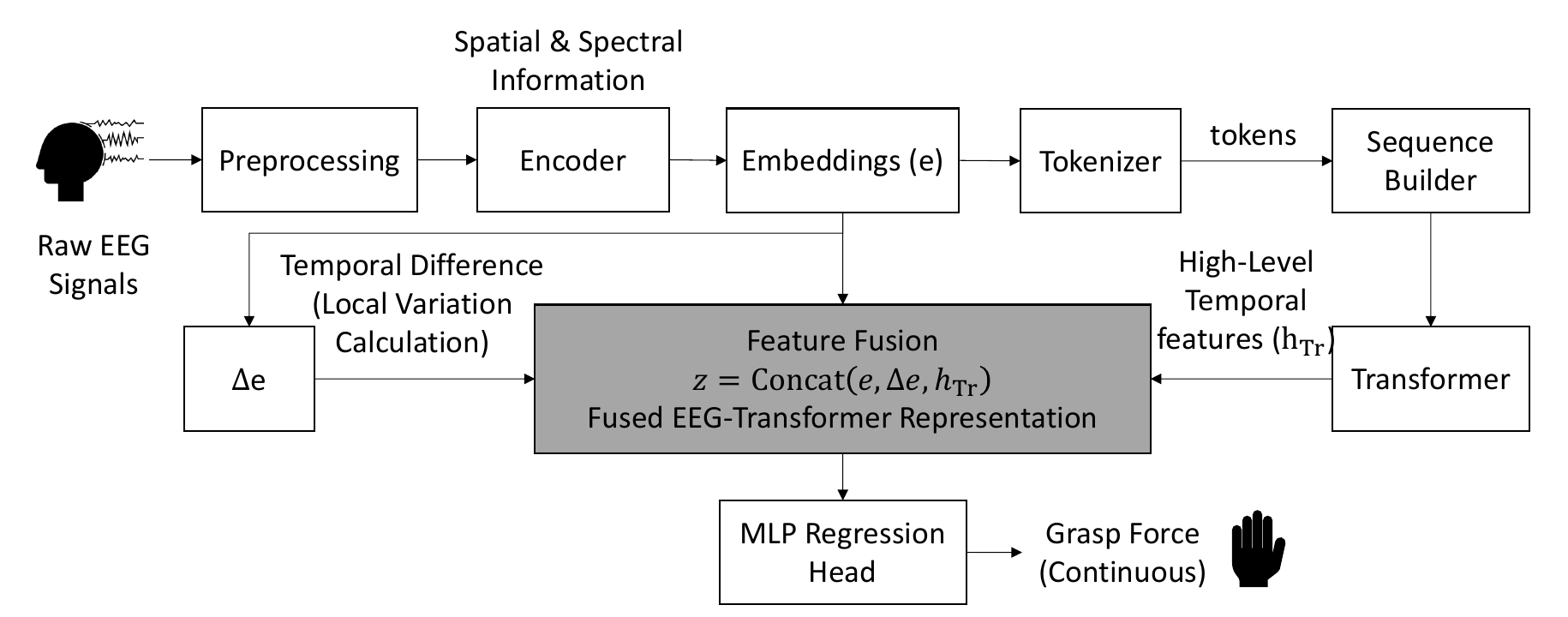}
    \caption{Proposed EEG-to-grasp-force regression framework. Raw EEG signals are first preprocessed and encoded using a CNN-GRU encoder to obtain embeddings ($e$). These embeddings are tokenised and organised into sequences for transformer-based temporal modelling. In parallel, local temporal variations ($\Delta e$) are computed. The final prediction is obtained by fusing embeddings ($e$), temporal differences ($\Delta e$), and transformer-derived features ($h_{\text{Tr}}$) through a regression head.}
    \label{fig:pipeline}
\end{figure*}

% We propose a tokenisation-based EEG decoding framework for continuous grasp-force prediction (Fig.~\ref{fig:pipeline}) comprising four stages: signal preprocessing and encoder-based embedding, discrete tokenisation, transformer-based temporal modelling, and fusion-based regression. 
We propose a tokenisation-based EEG decoding framework (see Fig.~\ref{fig:pipeline}) for continuous grasp-force prediction that integrates representation learning with transformer-based temporal modelling. The framework progresses through four stages: signal preprocessing and encoder-based embedding, discrete tokenisation, transformer-based temporal modelling, and fusion-based regression.

% The overall architecture is illustrated in Fig.~\ref{fig:pipeline}.

\subsection{Signal Preprocessing and Representation Learning}

The proposed framework is evaluated on the WAY-EEG-GAL dataset~\cite{luciw2014multi}, which contains multi-channel EEG recordings synchronised with continuous grasp-force measurements across multiple subjects performing object-interaction tasks. Raw EEG signals are preprocessed using standard procedures, including band-pass filtering (1--40~Hz), notch filtering at 50~Hz, and common average referencing~\cite{fatourechi2007emg}. The signals are normalised and clipped to stabilise training. A subset of motor-related channels (C3, C4, Cz, CP1, CP2, FC2, FC6) is selected to focus on task-relevant neural activity~\cite{ds09,Ehrsson2001-nf}.

The preprocessed EEG is segmented into fixed-length, overlapping, event-centred temporal windows capturing different phases of motor execution. An event-aware sampling strategy ensures balanced representation across motor states and transition dynamics. Each segment is represented as:
\begin{equation}
    X \in \mathbb{R}^{T \times C}, \quad y \in \mathbb{R}
\end{equation}
where $C$ is the number of channels and $T$ is the window length. The model learns the mapping:
\begin{equation}
    X_{t-\frac{T}{2}:t+\frac{T}{2}} \rightarrow y_t
\end{equation}
which captures both anticipatory and sustained neural activity without requiring explicit lag modelling~\cite{j94, Ehrsson2001-nf}.
Each EEG window is mapped to a compact embedding using a CNN-GRU encoder:
\begin{equation}
    e = f_{\text{enc}}(X), \quad e \in \mathbb{R}^{128}
\end{equation}
where convolutional layers extract spatial--spectral features and GRU layers model short-term temporal dynamics. Compared to LSTM-based architectures, this design reduces parameter complexity and improves generalisation~\cite{bouchane2025hybrid}.

The encoder is trained with a multi-objective loss that combines supervised force regression with representation regularisation; the contrastive term is inspired by~\cite{yue2022ts2vec}:
\begin{equation}
  \mathcal{L} =
  \alpha L_{\text{force}} +
  \beta L_{\text{feature}} +
  \gamma L_{\text{contrast}} +
  \delta L_{\text{temp}} +
  \eta L_{\text{var}}
\end{equation}
where $L_{\text{force}}$ supervises force prediction, $L_{\text{feature}}$ aligns embeddings with neurophysiological features, $L_{\text{contrast}}$ separates motor states, $L_{\text{temp}}$ enforces temporal consistency, and $L_{\text{var}}$ prevents representation collapse. The loss weights were set as $\alpha = 0.3$, $\beta = 0.2$, $\gamma = 0.05$, $\delta = 0.1$, and $\eta = 0.1$, determined empirically based on validation performance. GELU activation is used to improve stability for noisy EEG inputs~\cite{hendrycks2016gaussian}.

% \subsection{Discrete Tokenisation and Temporal Modelling}

% To enable sequence-based modelling, continuous EEG embeddings are converted into discrete tokens using a vector quantisation-based tokenizer~\cite{pradeepkumar2026tokenizing}. Given an embedding $e \in \mathbb{R}^{128}$, it is projected into a lower-dimensional space and discretised into finite levels:
% \begin{equation}
% z = f_{\text{proj}}(e), \quad z_q = \text{Quantize}(z)
% \end{equation}
% In practice, this is implemented using a lightweight neural projection followed by uniform scalar quantisation, resulting in a compact token representation. This transformation converts continuous features into symbolic sequences, enabling structured temporal modelling and compatibility with transformer-based learning. It also provides implicit regularisation against subject-specific embedding drift, reducing sensitivity to inter-subject variation~\cite{van2017neural}. The resulting token sequences are processed using a transformer to capture temporal dependencies~\cite{vaswani2017attention}:
% \begin{equation}
% h = f_{\text{tr}}(Z), \quad h \in \mathbb{R}^{128}
% \end{equation}
% where $Z$ denotes the input sequence (length $T=32$) and $h$ represents the learned temporal features. Tokens are projected into a higher-dimensional space, augmented with positional encoding, and then passed through stacked self-attention layers. During training, bidirectional attention is used, while a causal mask is applied during inference to ensure compatibility with streaming (simulated real-time) scenarios.
\subsection{Discrete Tokenisation and Temporal Modelling}

To enable sequence-based modelling, continuous EEG embeddings are converted into discrete tokens using a finite scalar quantisation (FSQ)-based tokeniser~\cite{pradeepkumar2026tokenizing}. Given an embedding $e \in \mathbb{R}^{128}$, it is projected into a lower-dimensional space and discretised into finite levels:
\begin{equation}
z = f_{\text{proj}}(e), \quad z_q = \text{Quantize}(z)
\end{equation}
In practice, this is implemented using a lightweight neural projection followed by uniform scalar quantisation, resulting in a compact token representation. This converts continuous features into symbolic sequences for structured temporal modelling, with implicit regularisation against subject-specific embedding drift~\cite{van2017neural}. FSQ is preferred over clustering-based alternatives (e.g., VQ-VAE) as it requires no learnable codebook, eliminates codebook collapse, and provides consistent subject-independent discretisation without per-subject calibration; essential for LOSO generalisation. The resulting token sequences are processed using a transformer to capture temporal dependencies~\cite{vaswani2017attention}:
\begin{equation}
h = f_{\text{tr}}(Z), \quad h \in \mathbb{R}^{128}
\end{equation}
where $Z$ denotes the input sequence (length $T=32$) and $h$ represents the learned temporal features. Tokens are projected into a higher-dimensional space, augmented with positional encoding, and then passed through stacked self-attention layers. During training, bidirectional attention is used, while a causal mask is applied during inference to ensure compatibility with streaming (simulated real-time) scenarios.
\subsection{Fusion-Based Prediction}

To integrate complementary representations, we employ a fusion-based regression framework that combines encoder embeddings, temporal differences, and transformer-derived features. The fused representation is defined as:
\begin{equation}
x = [h_{\text{Tr}}, e, \Delta e], \quad x \in \mathbb{R}^{384}
\end{equation}
where $h_{\text{Tr}} \in \mathbb{R}^{128}$ captures global temporal context, $e \in \mathbb{R}^{128}$ encodes spatial--spectral features, and $\Delta e \in \mathbb{R}^{128}$ represents local temporal variations between consecutive embeddings. The transformer is frozen as a fixed feature extractor during fusion training, and its outputs are aggregated using attention pooling~\cite{lin2017structured}. The fused representation is then processed using a multi-layer perceptron with GELU activations~\cite{hendrycks2016gaussian} to predict continuous grasp force.

\subsection{Evaluation Protocol}

The performance is evaluated on the WAY-EEG-GAL dataset~\cite{luciw2014multi} using a leave-one-subject-out (LOSO) cross-validation protocol, a standard strategy for evaluating subject-independent EEG decoding models~\cite{lotte2018review,dangi2025eegforcemap}. In this setting, 11 subjects are used for training and 1 for testing; the process is repeated across all 12 subjects, with results averaged across all evaluations. Two training settings are considered: (i) a pre-trained transformer with fold-wise fusion training, and (ii) strict LOSO, where both the transformer and fusion model are trained within each fold. In both settings, the encoder is trained on all twelve subjects, and only the downstream transformer and fusion stages are held out per fold, following the EEGForceMap convention~\cite{dangi2025eegforcemap}; the reported cross-subject results are therefore an upper-bound estimate of subject-independent performance. Performance is measured using the coefficient of determination~\cite{nagelkerke1991note}:
\begin{equation}
R^2 = 1 - \frac{\sum (y - \hat{y})^2}{\sum (y - \bar{y})^2}
\end{equation}

We consider two evaluation modes: offline and simulated real-time. In the offline setting, the model uses the full temporal window (past and future), enabling bidirectional processing. In the simulated real-time setting, only past context is used via a sliding buffer ($T = 32$) and causal masking~\cite{lotte2018review}. Experiments were conducted on a system with an NVIDIA RTX 3090 GPU, an Intel i7 CPU, and 32 GB RAM. Reported latency corresponds to inference time per sample under the streaming evaluation setup. Key hyperparameters: CNN-GRU encoder (embed dim 128, Adam, lr$=3{\times}10^{-4}$, 100 epochs); FSQ tokeniser ($K{=}8$ levels, token dim 24, 50 epochs); transformer ($d_\text{model}{=}128$, 4 heads, 4 layers, seq.\ length 32, AdamW, optimal 5 epochs); fusion MLP ($384{\to}[128,64]{\to}1$, Adam, 100 epochs).

%%%%%%%%%%%%%%%%%%%%%%%%%%%%%%%%%%%%%%%%%%%%%%%%%%
% RESULTS
%%%%%%%%%%%%%%%%%%%%%%%%%%%%%%%%%%%%%%%%%%%%%%%%%%

\section{Results}

\subsection{Temporal Modelling and Fusion Analysis}
\begin{figure*}[!t]
    \centering
    \includegraphics[width=0.40\textwidth]{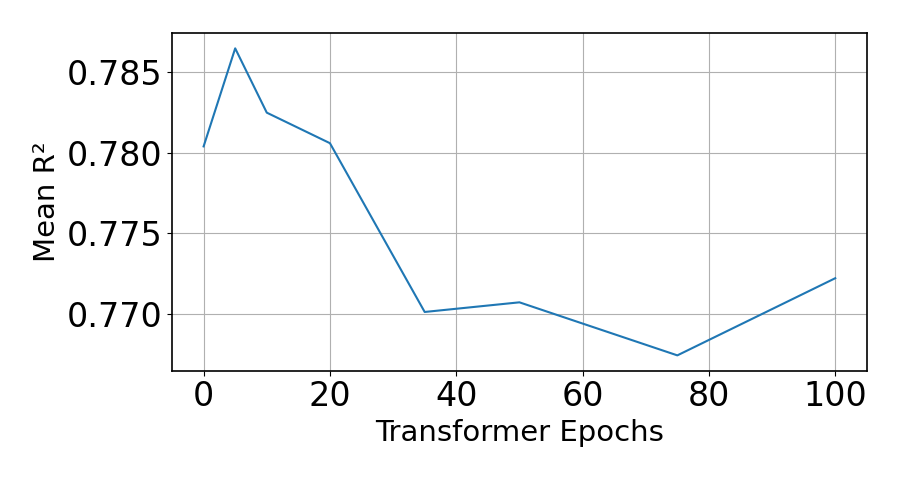}
    \includegraphics[width=0.36\textwidth]{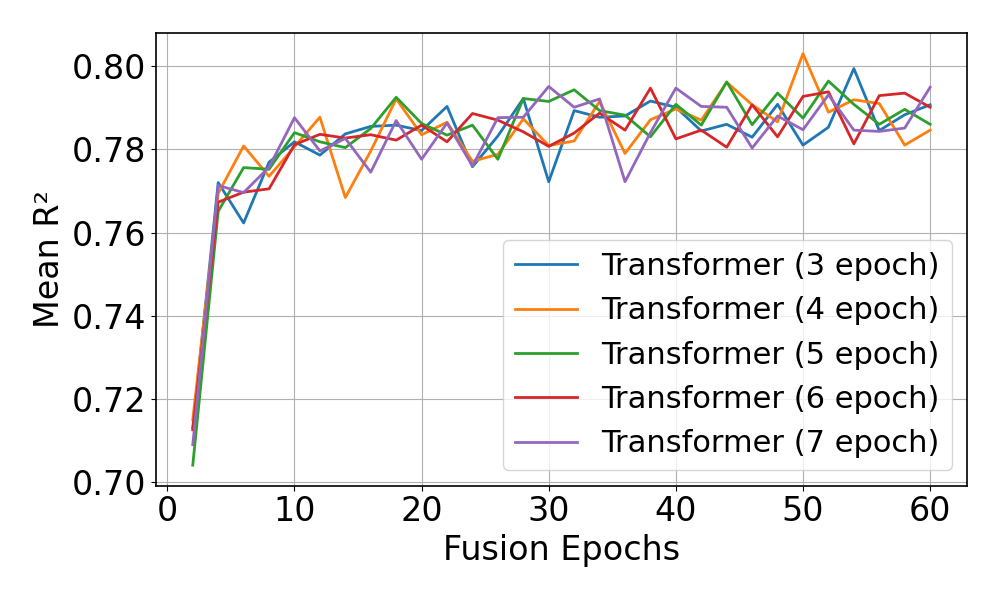}
    \caption{(Left) Effect of transformer training depth on LOSO performance. Performance peaks at approximately epoch~5. (Right) Fusion performance across transformer configurations (3--7 epochs).}
    \label{fig:llm_epoch}
    \label{fig:fusion_epoch}
\end{figure*}
We evaluate the impact of transformer training depth on LOSO performance. Performance peaks at early epochs (approximately 3--7 epochs), achieving $R^2 \approx 0.788$, and degrades with further training due to overfitting on subject-specific token distributions (Fig.~\ref{fig:llm_epoch})~\cite{kostas2021bendr}.

We further analyse the effect of fusion training across different transformer configurations (Fig.~\ref{fig:fusion_epoch}). All configurations reach their operating range ($R^2 \approx 0.78$) within the first 5--10 fusion epochs. Beyond this point, performance fluctuates within a narrow band ($\sigma \approx 0.008$), consistent with the stochastic nature of LOSO evaluation across 12 subjects rather than systematic learning trends. Transformer~(5 epochs) achieves the highest observed peak of $R^2 = 0.803$, motivating the selection of Tr(5)+F(100) as the reference configuration.

\subsection{Performance Evaluation}

\begin{table*}[!t]
    \centering
    \caption{Subject-wise performance across transformer--fusion configurations under LOSO evaluation. The narrow mean $R^2$ range (0.788--0.803) reflects framework stability rather than sensitivity to hyperparameter choice.}
    \label{tab:final_structured}
    \begin{tabular}{c|cc|cc|cc|cc|cc}
    % \scriptsize
        \toprule
        \multirow{2}{*}{Subject}
        & \multicolumn{2}{c|}{Tr(4) + F(50)}
        & \multicolumn{2}{c|}{Tr(7) + F(50)}
        & \multicolumn{2}{c|}{Tr(3) + F(75)}
        & \multicolumn{2}{c|}{Tr(5) + F(100)}
        & \multicolumn{2}{c}{Tr(6) + F(75)} \\
        
        % \cline{2-11}
        & $R^2$ & Lat
        & $R^2$ & Lat
        & $R^2$ & Lat
        & $R^2$ & Lat
        & $R^2$ & Lat \\
        
        \midrule
        
        P1  & 0.791 & 1.91 & 0.768 & 2.36 & 0.787 & 2.24 & 0.797 & 2.26 & 0.774 & 1.73 \\
        P2  & 0.760 & 2.06 & 0.788 & 1.95 & 0.765 & 2.15 & 0.787 & 2.36 & 0.791 & 2.60 \\
        P3  & 0.800 & 1.87 & 0.797 & 2.19 & 0.798 & 2.12 & 0.801 & 2.42 & 0.791 & 2.31 \\
        P4  & 0.776 & 1.92 & 0.768 & 2.17 & 0.776 & 2.31 & 0.787 & 2.22 & 0.789 & 1.92 \\
        P5  & 0.541 & 1.70 & 0.614 & 2.31 & 0.610 & 2.67 & 0.673 & 2.51 & 0.617 & 2.60 \\
        P6  & 0.817 & 1.90 & 0.834 & 2.92 & 0.838 & 2.17 & 0.821 & 2.24 & 0.841 & 2.18 \\
        P7  & 0.880 & 1.79 & 0.862 & 2.55 & 0.825 & 2.18 & 0.888 & 2.15 & 0.858 & 2.00 \\
        P8  & 0.795 & 2.20 & 0.795 & 2.20 & 0.794 & 2.14 & 0.802 & 2.22 & 0.795 & 1.72 \\
        P9  & 0.890 & 2.14 & 0.871 & 2.16 & 0.885 & 2.15 & 0.880 & 1.96 & 0.884 & 1.93 \\
        P10 & 0.823 & 1.86 & 0.812 & 2.36 & 0.830 & 2.23 & 0.800 & 2.42 & 0.826 & 2.25 \\
        P11 & 0.773 & 1.83 & 0.768 & 2.33 & 0.755 & 2.17 & 0.774 & 2.11 & 0.745 & 1.71 \\
        P12 & 0.813 & 1.72 & 0.832 & 2.28 & 0.829 & 2.33 & 0.824 & 2.05 & 0.835 & 1.81 \\
        
        \midrule
        
        Mean $R^2$
        & \multicolumn{2}{c|}{0.788}
        & \multicolumn{2}{c|}{0.793}
        & \multicolumn{2}{c|}{0.791}
        & \multicolumn{2}{c|}{0.803}
        & \multicolumn{2}{c}{0.795} \\
        
        Std $R^2$
        & \multicolumn{2}{c|}{0.084}
        & \multicolumn{2}{c|}{0.064}
        & \multicolumn{2}{c|}{0.065}
        & \multicolumn{2}{c|}{0.052}
        & \multicolumn{2}{c}{0.065} \\
        
        Mean Latency
        & \multicolumn{2}{c|}{1.91}
        & \multicolumn{2}{c|}{2.31}
        & \multicolumn{2}{c|}{2.24}
        & \multicolumn{2}{c|}{2.24}
        & \multicolumn{2}{c}{2.06} \\
        
        Max Latency
        & \multicolumn{2}{c|}{18.45}
        & \multicolumn{2}{c|}{18.82}
        & \multicolumn{2}{c|}{18.73}
        & \multicolumn{2}{c|}{17.36}
        & \multicolumn{2}{c}{20.97} \\
        
        \bottomrule
    \end{tabular}
\end{table*}
\begin{table*}[!t]
    \centering
    \caption{Subject-wise strict LOSO performance for the reference configuration Tr(5)+F(100).}
    \label{tab:latency_subject}
    \scriptsize
    \begin{tabular}{l|cccccccccccc|cc}
        \toprule
        & \textbf{P1} & \textbf{P2} & \textbf{P3} & \textbf{P4} & \textbf{P5} & \textbf{P6} & \textbf{P7} & \textbf{P8} & \textbf{P9} & \textbf{P10} & \textbf{P11} & \textbf{P12} & \textbf{Mean} & \textbf{Std} \\
        \midrule
        $R^2$        & 0.786 & 0.782 & 0.800 & 0.785 & 0.647 & 0.801 & 0.833 & 0.792 & 0.875 & 0.827 & 0.750 & 0.836 & \textbf{0.793} & \textbf{0.054} \\
        Latency (ms) & 1.70  & 1.66  & 1.68  & 1.71  & 1.57  & 1.60  & 1.59  & 1.69  & 1.67  & 1.67  & 1.71  & 1.65  & \textbf{1.66}  & --             \\
        \bottomrule
    \end{tabular}
\end{table*}
% \begin{table}[!htbp]
%     \centering
%     \caption{Subject-wise strict LOSO performance for the reference configuration Tr(5)+F(100).}
%     \label{tab:latency_subject}
%     % \scriptsize
%     \begin{tabular}{lcc}
%         \toprule
%         \textbf{Subject} & \textbf{$R^2$} & \textbf{Latency (ms)} \\
%         \midrule
%         P1  & 0.786 & 1.70 \\
%         P2  & 0.782 & 1.66 \\
%         P3  & 0.800 & 1.68 \\
%         P4  & 0.785 & 1.71 \\
%         P5  & 0.647 & 1.57 \\
%         P6  & 0.801 & 1.60 \\
%         P7  & 0.833 & 1.59 \\
%         P8  & 0.792 & 1.69 \\
%         P9  & 0.875 & 1.67 \\
%         P10 & 0.827 & 1.67 \\
%         P11 & 0.750 & 1.71 \\
%         P12 & 0.836 & 1.65 \\
%         \midrule
%         \textbf{Mean} & \textbf{0.793} & \textbf{1.66} \\
%         \textbf{Std}  & \textbf{0.054} & -- \\
%         \bottomrule
%     \end{tabular}
% \end{table}

\begin{figure*}[!t]
    \centering
    \includegraphics[width=0.30\textwidth]{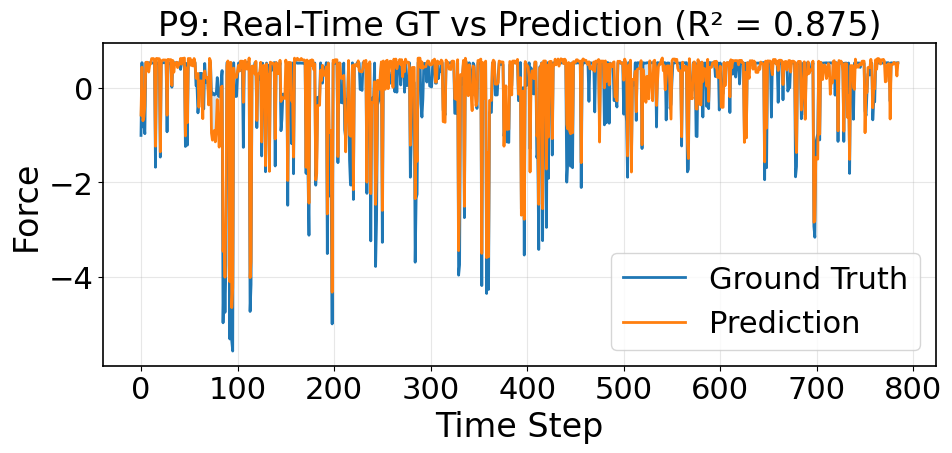}
    \includegraphics[width=0.30\textwidth]{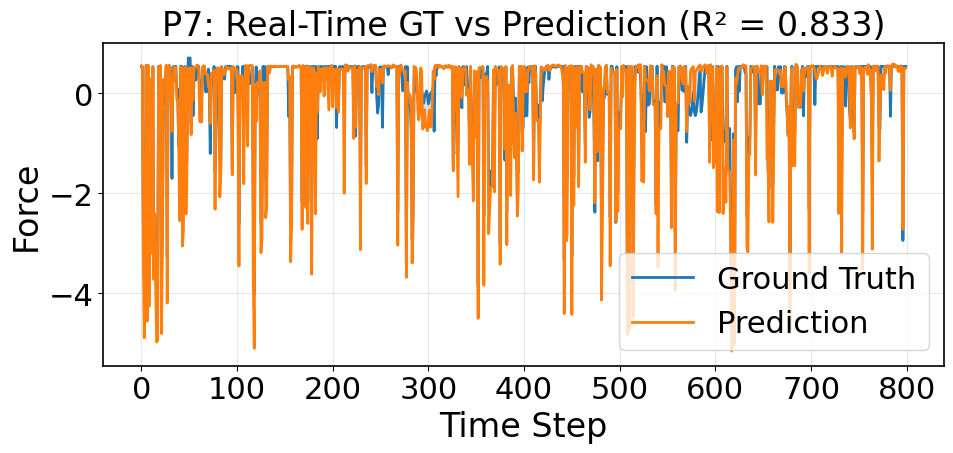}
    \includegraphics[width=0.30\textwidth]{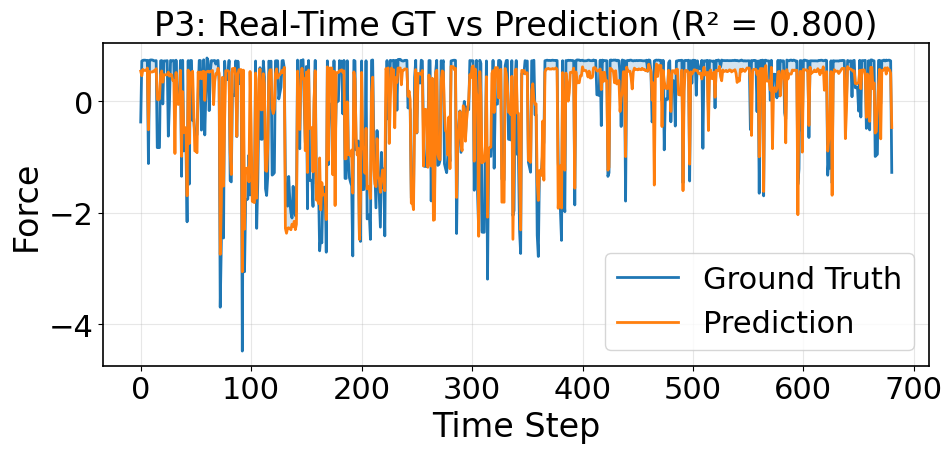}
    \includegraphics[width=0.30\textwidth]{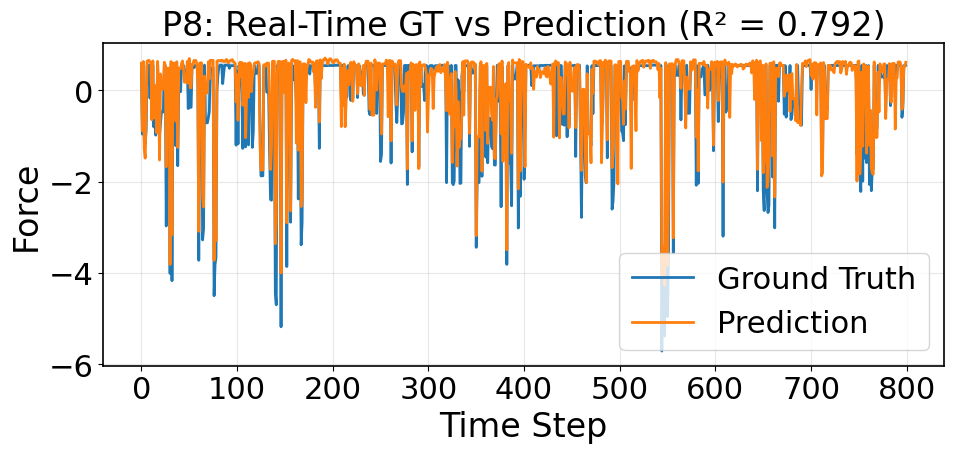}
    \includegraphics[width=0.30\textwidth]{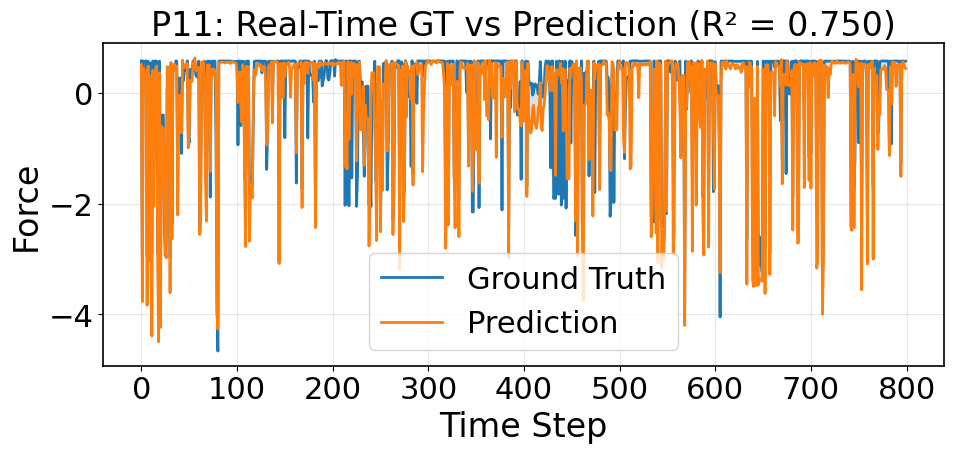}
    \includegraphics[width=0.30\textwidth]{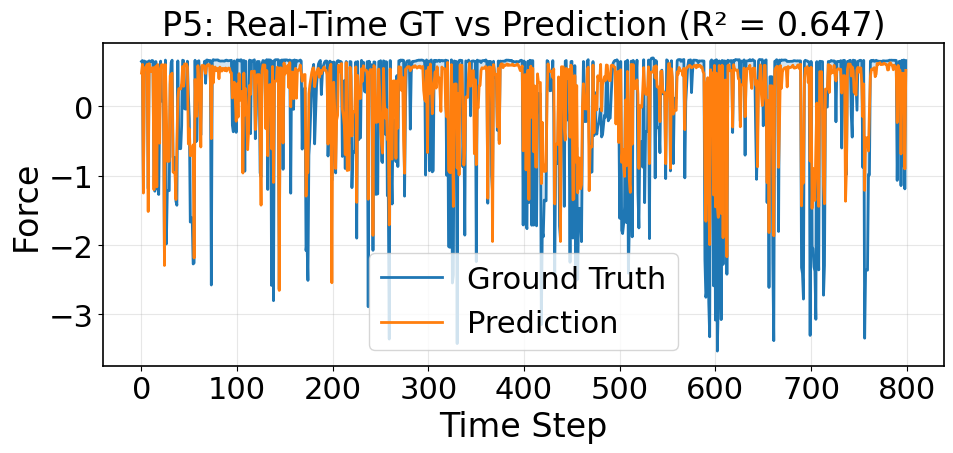}
    \caption{Representative real-time predictions for six subjects spanning the observed performance range: best (P9, P7), average (P3, P8), and worst (P11, P5). Each subplot shows predicted vs.\ ground-truth force over an 800-sample segment capturing force onset, sustain, and release. Full per-subject results are reported in Table~\ref{tab:latency_subject}.}
    \label{fig:qualitative}
\end{figure*}

We evaluate the proposed framework in simulated real-time and cross-subject settings using LOSO. Across all configurations, mean $R^2$ ranges narrowly from 0.788 to 0.803 (Table~\ref{tab:final_structured}), consistent with the rapid convergence and stable plateau observed in Fig.~\ref{fig:fusion_epoch}. Tr(5)+F(100) achieves the highest mean $R^2 = 0.803 \pm 0.052$ and is used as the reference configuration. The overlapping standard deviations across all configurations (mean range 0.015, well within the reported across-subject standard deviations of 0.052--0.084) indicate that performance differences reflect stochastic variation under LOSO rather than systematic effects of hyperparameter choice, demonstrating that the framework is robust to both transformer depth and fusion training duration (Friedman test: $\chi^2(4){=}4.48$, $p{=}0.345$; no significant difference across configurations).

The model maintains low inference latency across configurations (approximately 1.9--2.3~ms per sample; Table~\ref{tab:final_structured}). The best-performing configuration achieves low latency (2.24~ms), confirming that improved predictive performance does not introduce additional computational overhead. The higher maximum latency values correspond to occasional initialisation or buffering overhead during streaming inference, while steady-state latency remains consistently low across samples. Under strict LOSO evaluation, the model achieves a mean $R^2 = 0.793 \pm 0.054$ (Table~\ref{tab:latency_subject}). Most subjects achieve $R^2 > 0.75$, with several exceeding 0.80, indicating strong cross-subject generalisation. Lower performance for P5 ($R^2 = 0.647$) is discussed in Section~\ref{sec:discussion}.

Qualitative analysis (Fig.~\ref{fig:qualitative}) shows that the predicted signals closely follow ground-truth trends, accurately capturing key temporal transitions, including force onset, sustain, and release. Minor discrepancies in high-amplitude regions are observed, consistent with the noisy nature of EEG signals.

\begin{table*}[!t]
    \centering
    \caption{Comparison with prior methods for EEG-based grasp force decoding. SS = subject-specific; SI = subject-independent (LOSO). Best SI result in \textbf{bold}.}
    \label{tab:Comparison}
    % \scriptsize
    \begin{tabular}{lccccc}
        \toprule
        \textbf{Models} & \textbf{Model Type} & \textbf{Data Type} & \textbf{Decoding Approach} & \multicolumn{2}{c}{\textbf{CoD ($R^2$)}} \\
        \cmidrule(lr){5-6}
         &  &  &  & \textbf{SS} & \textbf{SI} \\
        \midrule
        
        Multi-Modal Decoder~\cite{fu2023eeg} & CNN-LSTM & EEG + EMG & Deep Learning & 0.457 & -- \\
        Classifier-Decoder Ensemble~\cite{pgpsc19} & LDA + CNN-LSTM & EEG & Ensemble & 0.487 & -- \\
        Brain-Inspired Spike Neural Network~\cite{kumarasinghe2021brain} & SNN & EEG & Deep Learning & 0.504 & -- \\
        
        \midrule
        EEGForceMap + SFLR~\cite{dangi2025eegforcemap} & LR & EEG & Linear & 0.261 & 0.156 \\
        EEGForceMap + MLR~\cite{dangi2025eegforcemap} & MLR & EEG & Linear & 0.425 & 0.356 \\
        EEGForceMap + PLSR~\cite{dangi2025eegforcemap} & PLSR & EEG & Non-linear & 0.536 & 0.405 \\
        EEGForceMap + NNR~\cite{dangi2025eegforcemap} & MLP & EEG & Deep Learning & 0.815 & 0.785 \\
        
        \midrule
        \textbf{Proposed (Offline LOSO)} & Transformer + Fusion & EEG & Deep Learning & -- & \textbf{0.817} \\
        \textbf{Proposed (Simulated Real-Time LOSO)} & Transformer + Fusion & EEG & Deep Learning & -- & \textbf{0.793} \\
        
        \bottomrule
    \end{tabular}
\end{table*}

\subsection{Comparison with Prior Work}

We compare the proposed approach with prior methods for EEG-based force decoding, including the EEGForceMap pipeline~\cite{dangi2025eegforcemap}. Table~\ref{tab:Comparison} summarises the coefficient of determination ($R^2$) under subject-specific (SS) and subject-independent (SI/LOSO) settings. The strongest baseline, EEGForceMap + NNR~\cite{dangi2025eegforcemap}, achieves $R^2$ = 0.785 in the subject-independent setting. The proposed method achieves $R^2$ = 0.817 in offline LOSO and $R^2$ = 0.793 $\pm$ 0.054 in simulated real-time LOSO (Table~\ref{tab:Comparison}), surpassing the best baseline by 3.2\% in offline settings under identical LOSO evaluation (10 of 12 subjects individually exceeded NNR; binomial sign test, $p{=}0.019$). Unlike prior approaches that rely on handcrafted features, the proposed framework learns both continuous and tokenised representations directly from EEG signals, enabling effective modelling of temporal dependencies while maintaining cross-subject generalisation.

% \subsection{Ablation Study}

% To better understand the contribution of individual components, we perform an ablation study (Table~\ref{tab:ablation}).

% \begin{table}[!t]
%     \centering
%     \caption{Ablation study of different components in the proposed framework}
%     \label{tab:ablation}
%     % \scriptsize
%     \begin{tabular}{l c}
%         \toprule
%         \textbf{Model Configuration} & \textbf{R$^2$ (LOSO)} \\
%         \midrule
%         Encoder (Embedding only) & $\sim$0.64 \\
%         Embedding + Transformer & $\sim$0.78 \\
%         Token + Transformer & $\sim$0.70 \\
%         \textbf{Fusion (Embedding + Token + Transformer)} & \textbf{0.817} \\
%         \bottomrule
%     \end{tabular}
% \end{table}

% Using only encoder embeddings yields moderate performance ($R^2 \sim 0.64$), indicating that continuous representations capture useful spatial--spectral information but lack sufficient temporal modelling. Tokenised representations with transformer-based modelling improve temporal structure ($R^2 \sim 0.70$), but suffer from reduced accuracy due to the loss of fine-grained signal detail. Combining continuous embeddings with transformer-based temporal modelling recovers much of this fidelity ($R^2 \sim 0.78$). The best performance is achieved using the full fusion model ($R^2 = 0.817$), confirming that no single representation type is sufficient and that their combination is necessary for accurate force decoding.

\subsection{Ablation Study}
To understand the contribution of each component, we conduct an ablation study, with the results presented in Table~\ref{tab:ablation}. Encoder embeddings alone already achieve strong performance ($R^2 \sim 0.81$), confirming that the multi-task encoder captures rich force-relevant spatial--spectral information. Applying a supervised transformer to continuous embedding sequences ($R^2 \sim 0.78$) or to discrete token sequences ($R^2 \sim 0.76$) does not exceed this, indicating that temporal modelling over a single representation is prone to inter-subject overfitting under LOSO. The full fusion ($R^2 = 0.817$) is comparable to the encoder-only baseline within LOSO variance: tokenisation and transformer-based temporal modelling preserve decoding accuracy while contributing the compact, frozen, streaming representation used for real-time inference, rather than additional single-window accuracy.
\begin{table}[!t]
    \centering
    \caption{Summary of ablation study findings.}
    \label{tab:ablation}
    % \scriptsize
    \begin{tabular}{l c}
        \toprule
        \textbf{Model Configuration} & \textbf{R$^2$ (LOSO)} \\
        \midrule
        Encoder (Embedding only) & $\sim$0.81 \\
        Embedding + Transformer & $\sim$0.78 \\
        Token + Transformer & $\sim$0.76 \\
        \textbf{Fusion (Embedding + Token + Transformer)} & \textbf{0.817} \\
        \bottomrule
    \end{tabular}
\end{table}

%%%%%%%%%%%%%%%%%%%%%%%%%%%%%%%%%%%%%%%%%%%%%%%%%%
% DISCUSSION
%%%%%%%%%%%%%%%%%%%%%%%%%%%%%%%%%%%%%%%%%%%%%%%%%%

\section{Discussion}
\label{sec:discussion}
\textit{Why tokenisation: a frozen, reusable temporal backbone:} A central design goal is a temporal model pretrained once and reused, frozen, for real-time prediction across subjects. Tokenisation provides the required stable, shared vocabulary: quantising embeddings into a finite FSQ codebook yields a subject-invariant token space over which a causal transformer is pretrained self-supervised and then frozen, producing temporal context in $\approx$2~ms per window with no per-subject retraining. The ablation (Table~\ref{tab:ablation}) shows this branch preserves decoding accuracy rather than increasing it; its contribution is architectural, analogous to how a fixed sub-word vocabulary underpins a pretrained language model. With only twelve subjects, the token transformer cannot learn rich sequential statistics; larger EEG corpora may enable transferable sequence priors that a per-window encoder cannot, a primary direction of future work.

\textit{Transformer as a regularised temporal feature extractor:}
Transformers are widely used for sequence modelling due to their ability to capture long-range dependencies~\cite{vaswani2017attention}, but are prone to overfitting in limited-data EEG settings~\cite{kostas2021bendr}. As shown in Fig.~\ref{fig:llm_epoch}, performance peaks at approximately epoch~5 ($R^2 \approx 0.788$) and degrades monotonically with deeper optimisation, as the transformer begins memorising subject-specific token distributions that do not transfer. Freezing the transformer after shallow training and optimising only the fusion MLP decouples temporal feature extraction from regression. As Fig.~\ref{fig:fusion_epoch} confirms, all configurations converge within 10 fusion epochs, and the epoch-wise variance ($\sigma \approx 0.008$) reflects LOSO stochasticity rather than continued learning, validating the robustness of this two-stage training strategy.

\textit{Robustness and generalisation in simulated real-time:}
In simulated real-time evaluation, the discrete tokenisation step provides implicit regularisation against embedding drift across the sliding buffer window, complementing the amplitude-preserving continuous features during sequential prediction (Table~\ref{tab:latency_subject}, Fig.~\ref{fig:qualitative}). The proposed framework achieves strong cross-subject performance ($R^2 = 0.793 \pm 0.054$) with low inference latency ($\approx$1.6--2~ms per sample), confirming practical viability for real-time EEG decoding.

\textit{Learning dynamics and qualitative analysis:}
Qualitative analysis (Fig.~\ref{fig:qualitative}) shows accurate tracking of force onset, plateau, and release across the full performance range, indicating sensitivity to temporal motor structure rather than amplitude heuristics. Residual errors concentrate during high-amplitude, high-noise episodes, consistent with the limitations of scalp EEG. The reduced performance of P5 ($R^2 = 0.647$) reflects BCI illiteracy rather than a systematic model failure, as all other subjects achieve $R^2 \geq 0.75$.

% \paragraph{Limitations and future directions}
% Evaluation is limited to a single dataset, and further validation across diverse EEG datasets is required to assess generalisability. Practical deployment in closed-loop BCI systems introduces additional challenges beyond what is captured here, including feedback delay, signal drift, and user adaptation. Future work should investigate these aspects in real-world settings. A promising direction is the development of richer tokenisation strategies that capture higher-level structure in neural signals, such as learned vector-quantised codebooks or hierarchical discretisation, which could better align EEG representations with pre-trained neural foundation models. 
\textit{Limitations and future directions:}
Evaluation is limited to the WAY-EEG-GAL dataset~\cite{luciw2014multi}, and further validation across diverse EEG datasets is required to assess generalisability. The reported cross-subject scores are also an upper bound, as the encoder is trained on all subjects; fully subject-independent representation learning remains an open challenge. Practical deployment in closed-loop BCI systems introduces additional challenges beyond what is captured here, including
feedback delay, signal drift, and user adaptation. A promising direction is the development of richer tokenisation
strategies that capture higher-level structure in neural signals, such as learned VQ codebooks or hierarchical discretisation, which could better align
EEG representations with pre-trained neural foundation models; direct interpretability of the token space with respect to motor-state structure remains an open research direction.
%%%%%%%%%%%%%%%%%%%%%%%%%%%%%%%%%%%%%%%%%%%%%%%%%%
% CONCLUSION
%%%%%%%%%%%%%%%%%%%%%%%%%%%%%%%%%%%%%%%%%%%%%%%%%%

\section{Conclusion}

We presented EEGForceFusion, a hybrid EEG decoding framework that jointly models continuous and tokenised representations for subject-independent grasp force prediction. Evaluated under strict leave-one-subject-out conditions on the WAY-EEG-GAL dataset, the framework achieves $R^2$ = 0.817 in offline settings and $R^2$ = 0.793 in simulated real-time evaluation, surpassing the strongest baseline by 3.2\%, with inference latency of $\approx$1.6--2~ms per sample. A central finding is that the transformer is most effective as a shallow feature extractor: decoupling its training from the fusion regression head prevents subject-specific overfitting and yields a framework robust to both transformer depth and fusion training duration. The ablation study shows that fusing tokenised and continuous representations matches encoder-only accuracy while providing a frozen, reusable temporal backbone suited to real-time deployment. Representing neural signals as structured token sequences offers a promising bridge between raw EEG and modern transformer-based architectures, with potential impact in assistive robotics, neuro-rehabilitation, and next-generation BCI systems. All source code and dataset details are available at the EEGForceFusion repository\footnote{\url{https://github.com/HAIx-Lab/EEGForceFusion}} to ensure reproducibility. 

\noindent\textbf{Acknowledgement:} This study was supported by the IIT Gandhinagar startup grant (IP/IITGN/CSE/YM/2324/05) and the IndiaAI Fellowship (PG251215005117).

%%%%%%%%%%%%%%%%%%%%%%%%%%%%%%%%%%%%%%%%%%%%%%%%%%
% \bibliographystyle{IEEEtran}
\bibliography{references}

\end{document}